\documentclass[letterpaper, 10 pt, journal, twoside]{IEEEtran}  

\IEEEoverridecommandlockouts                              

\title{
Control-guided Communication:\\ Efficient Resource Arbitration and Allocation in Multi-hop Wireless Control Systems
}%

\author{Dominik Baumann$^{1,*}$, Fabian Mager$^{2,*}$, Marco Zimmerling$^2$, and Sebastian Trimpe$^{1}$
\thanks{$^{*}$Equal contribution.}%
\thanks{$^{1}$Intelligent Control Systems Group, Max Planck Institute for Intelligent Systems, Stuttgart/T\"{u}bingen, Germany. 
Email: dbaumann@tuebingen.mpg.de, trimpe@is.mpg.de.}%
\thanks{$^{2}$Networked Embedded Systems Lab, TU Dresden, Dresden, Germany.
Email: \{fabian.mager, marco.zimmerling\}@tu-dresden.de}%
\thanks{This work was supported in part by the German Research
Foundation within the Cluster of Excellence cfaed (grant 
EXC 1056), SPP 1914 (grants ZI 1635/1-1 and TR 1433/1-1), and 
the Emmy Noether project NextIoT (grant ZI 1635/2-1); the Cyber
Valley Initiative; and the Max Planck Society.}
} 

\usepackage[english]{babel}
\usepackage{inputenc}
\usepackage{balance}
\usepackage{xspace}
\usepackage{rotating}
\usepackage{url}
\usepackage{cite}
\usepackage{csquotes}
\usepackage{subcaption}
\usepackage{paralist}
\usepackage{fixltx2e}
\usepackage{multirow}
\usepackage{amsmath,amssymb,mathtools}
\usepackage{xfrac}
\usepackage{array}
\usepackage{textcomp}
\usepackage{siunitx}
\usepackage{microtype}
\usepackage[dvipsnames]{xcolor}
\usepackage{pgfplots}
\pgfplotsset{compat=newest,unit code/.code={\si{#1}},plot coordinates/math parser=false,grid style={lightgray}}
\usepgfplotslibrary{units,external,groupplots}
\usetikzlibrary{positioning,angles,quotes,patterns,shapes}

\usepackage[left=52pt, right=52pt, bottom=50pt, top=60pt]{geometry}

\graphicspath{{./figures/}}

\DeclareSIUnit{\belmilliwatt}{Bm}
\DeclareSIUnit{\dBm}{\deci\belmilliwatt}

\DeclareMathOperator*{\E}{\mathbb{E}}
\DeclareMathOperator*{\Var}{Var}
\DeclareMathOperator*{\Tr}{Tr}
\newcommand{\norm}[1]{\left\lVert#1\right\rVert}

\DeclareMathOperator*{\R}{\mathbb{R}}

\newcommand{\transp}{\text{T}}

\let\originalleft\left
\let\originalright\right
\renewcommand{\left}{\mathopen{}\mathclose\bgroup\originalleft}
\renewcommand{\right}{\aftergroup\egroup\originalright}

\newcommand\figref[1]{Fig.~\ref{#1}}
\newcommand\tabref[1]{Table~\ref{#1}}
\newcommand\secref[1]{Sec.~\ref{#1}}

\newcommand{\eg}{\emph{e.g.},\xspace}
\newcommand{\ie}{\emph{i.e.},\xspace}
\newcommand{\etc}{etc.\xspace}

\usepackage{ifthen}
\newboolean{authnotes}

\setboolean{authnotes}{false}

\ifthenelse{\boolean{authnotes}}
{
\newcommand{\fm}[1]{\footnote{{\bf\color{blue} Fabian: #1}}}
\newcommand{\mz}[1]{\footnote{{\bf\color{blue} Marco: #1}}}
\newcommand{\db}[1]{\footnote{{\bf\color{green!50!black} Dominik: #1}}}
\newcommand{\st}[1]{\footnote{{\bf\color{purple!90!black} Sebastian: #1}}}
\newcommand{\rj}[1]{\footnote{{\bf\color{orange!50!black} Romain: #1}}}

}
{
\newcommand{\fm}[1]{}
\newcommand{\mz}[1]{}
\newcommand{\db}[1]{}
\newcommand{\st}[1]{}
\newcommand{\rj}[1]{}

}

\tikzstyle{block} = [draw, rectangle, minimum height=2em, minimum width=3em]
\tikzstyle{addon} = [draw, rectangle, rounded corners]
\tikzstyle{pinstyle} = [pin edge={<-,thin,black}]
\tikzstyle{pinstyle2} = [pin edge={->,thin,black}]
\tikzstyle{mult} = [draw, isosceles triangle]
\tikzstyle{circ} = [draw, circle]
\tikzstyle{coord} = [coordinate]
\tikzstyle{circ2} = [draw, circle,minimum width=3pt, inner sep=0]
\tikzset{>=latex}
\tikzset{radiation/.style={{decorate,decoration={expanding
waves,angle=90,segment length=4pt}}}}
\usepackage{booktabs}
\usepackage{pifont}
\newcommand{\cmark}{\textcolor{green!50!black}{\ding{51}}}%
\newcommand{\xmark}{\textcolor{red}{\ding{55}}}%

\definecolor{dark-gray}{gray}{0.35}

\newcommand{\dpp}{DPP\xspace}

\newcommand{\fakepar}[1]{\vspace{0mm}\noindent\textbf{#1.}}

\usepackage{fancyhdr}
\newcommand{\mytitle}{\textbf{Accepted final version.}
To appear in \textit{IEEE Control Systems Letters}.\\
\copyright 2019 IEEE. Personal use of this material is permitted. Permission
from IEEE must be obtained for all other uses, in any current or future
media, including reprinting/republishing this material for advertising or
promotional purposes, creating new collective works, for resale or
redistribution to servers or lists, or reuse of any copyrighted component of
this work in other works.}
\begin{document}

\maketitle
\thispagestyle{fancy}	
\pagestyle{empty}

%
\begin{abstract}
In future autonomous systems, wireless multi-hop communication is key to enable collaboration among distributed agents at low cost and high flexibility.
When many agents need to transmit information over the same wireless network, communication becomes a shared and contested resource.
Event-triggered and self-triggered control account for this by transmitting data only when needed, enabling significant energy savings.
However, a solution that brings those benefits to multi-hop networks \emph{and} can reallocate freed up bandwidth to additional agents or data sources is still missing.
To fill this gap, we propose \emph{control-guided communication}, a novel co-design approach for distributed self-triggered control over wireless multi-hop networks. 
The control system informs the communication system of its transmission demands ahead of time, and the communication system allocates resources accordingly.
Experiments on a cyber-physical testbed show that multiple cart-poles can be synchronized over wireless, while serving other traffic when resources are available, or saving energy.
These experiments are the first to demonstrate and evaluate distributed self-triggered control over low-power multi-hop wireless networks at update rates of tens of milliseconds.%
\end{abstract}
\begin{IEEEkeywords}
Wireless control systems, self-triggered control.
\end{IEEEkeywords}
\vspace{-5mm}


\section{Introduction}
\label{sec:intro}

\IEEEPARstart{T}{he} unparalleled flexibility and cost efficiency when closing feedback loops over wireless networks enables many cyber-physical applications.
For instance, in a smart factory, plants are controlled via remote controllers, mobile robots interact with the plants, and distributed sensors provide additional measurements.
Another example is drones regularly exchanging data to fly in formation. 
These and other applications demand wireless \emph{multi-hop} communication to cover large distances and \emph{fast} update intervals of tens of milliseconds to keep up with the dynamics of the systems to be controlled~\cite{Akerberg2011}.

\fakepar{Challenges}
Fast feedback control over wireless multi-hop networks is challenging owing to the inherent imperfections of wireless networks, such as transmission delays and message loss.
Moreover, the limited network bandwidth can lead to congestion when many agents need to communicate at the same time, and wireless radios draw considerable power, which is a major concern for embedded sensors and mobile devices that must be untethered and thus powered by batteries.
For these reasons, adaptive schemes are needed where agents use the network only when necessary to \emph{save energy}, and available resources are \emph{reallocated} at run time to serve those in need.

To use the limited bandwidth and energy more efficiently, event-triggered control~(ETC) and self-triggered control~(STC) methods have been developed~\cite{HeJoTa12,Mi15}.
Unlike periodic control, in ETC and STC the decision whether to communicate or not is based on events, such as an error exceeding a threshold.
ETC \emph{instantaneously} decides whether to communicate, leaving no time to save energy or reallocate bandwidth in case of a negative triggering decision.
STC, instead, decides \emph{ahead of time} about the next triggering instant.
However, to utilize freed resources (\eg to serve traffic from additional remote sensors), an integration of STC designs and wireless communication protocols is required.
Moreover, such co-design approaches must be evaluated on real cyber-physical testbeds to establish trust in feedback control over wireless~\cite{lu2016real}.
While a large body of work on STC exists (see \cite{HeJoTa12,Mi15,VeMaFu03,WaLe09,MaAnTa10} and the references therein), the integration of STC designs with wireless protocols including an experimental evaluation has rarely been considered.
\mbox{The few exceptions are listed in \tabref{tab:rel_work} and discussed next.}

\begin{table}[!tb]
\caption{Qualitative comparison of prior and our work on integrating STC with wireless communication, evaluated through real-world experiments.}
\vspace{-1mm}
\label{tab:rel_work}
\begin{tabular}{cccccc}
\toprule
\multirow{2}{*}{Work}  & Fast update & Multi- & Energy & Reallo- & Distributed\\
& intervals & hop & savings & cation & implementation\\
\midrule
\cite{Araujo2014}  & \xmark & \xmark & \cmark & \cmark & \xmark\\
\cite{araujo2011self} & \xmark & \xmark & \cmark & \cmark & \xmark \\
\cite{ma2018efficient}  & \xmark & \cmark & \cmark & \xmark & \xmark \\
\cite{santos2014adaptive}  & \cmark & \xmark & \cmark & \xmark & \xmark \\
\cite{santos2015aperiodic}  & \cmark & \xmark & \cmark & \xmark & \xmark \\
\textbf{This} & \cmark & \cmark & \cmark & \cmark & \cmark \\
\bottomrule
\end{tabular}
\vspace{-5mm}
\end{table}

%

\fakepar{Prior work}
Existing approaches integrating STC and wireless communication target remote control, for example, of a double-tank process~\cite{Araujo2014,araujo2011self}, a simulated load-positioning system~\cite{ma2018efficient}, or a mobile robot~\cite{santos2014adaptive}.
Coordination in multi-robot systems has been studied in~\cite{santos2015aperiodic}, but the control commands are computed by a central entity, so the implementation is not distributed.
All works show that STC allows for solving the control task with less communication than periodic control, enabling significant energy savings. 
However, reallocation of freed resources has only been demonstrated 
in~\cite{araujo2011self,Araujo2014}, for single-hop networks and update intervals of a few seconds.
In fact, STC over a wireless multi-hop network has only been shown in~\cite{ma2018efficient}, with an update interval of \SI{1}{\second}.

In summary, no solution exists that provides energy savings \emph{and} reallocation of freed resources for the control of systems at fast update intervals over multi-hop networks.
Moreover, no work has shown a distributed implementation of a STC law, where agents locally use information obtained over the network to solve a common control task.
However, a complete solution is needed to enable novel applications, such as collaborative multi-robot swarms for future smart production systems. 

\fakepar{Contribution}
We present a co-design of control and communication for multi-hop wireless networks that fills this gap.
Our approach arbitrates the available communication bandwidth among different types of traffic from any entity in the network, while simultaneously shutting down resources completely to save energy when neither the control system nor any other entity needs the full bandwidth.
We evaluate the approach on a three-hop cyber-physical testbed with multiple physical systems~\cite{Baumann2018}, demonstrating improved resource efficiency at high control performance for update intervals below \SI{100}{\milli\second}.

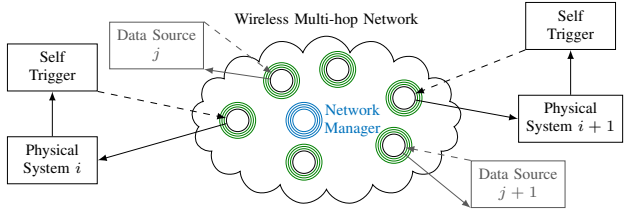
\begin{figure}
\centering
\resizebox{0.48\textwidth}{!}{
\tikzset{font=\scriptsize}
\tikzset{>=latex}
\begin{tikzpicture}
\node[ align = center ,draw , text = black, cloud ,cloud puffs =17 , cloud puff arc =140 ,  aspect =2.5,fill=white,minimum width = 12em,minimum height = 7.5em,label=above:{Wireless Multi-hop Network}](nw){};
\node[draw,circle,minimum size = 10pt]at([yshift=2em,xshift = 3em]nw.south)(start1){};
\node[draw,circle,minimum size = 12pt, green!50!black]at(start1.center){};
\node[draw,circle,minimum size = 14pt, green!50!black]at(start1.center){};
\node[draw,circle,minimum size = 16pt, green!50!black]at(start1.center){};
\node[draw,circle,minimum size = 10pt]at([yshift=-2em,xshift = -2em]nw.north)(start2){};
\node[draw,circle,minimum size = 12pt, green!50!black]at(start2.center){};
\node[draw,circle,minimum size = 14pt, green!50!black]at(start2.center){};
\node[draw,circle,minimum size = 16pt, green!50!black]at(start2.center){};

\node[draw,circle,minimum size = 10pt]at([xshift=2em]nw.west)(start3){};
\node[draw,circle,minimum size = 12pt, green!50!black]at(start3.center){};
\node[draw,circle,minimum size = 14pt, green!50!black]at(start3.center){};
\node[draw,circle,minimum size = 16pt, green!50!black]at(start3.center){};

\node[draw,circle,minimum size = 10pt]at([xshift=-2.5em,yshift=1em]nw.east)(start4){};
\node[draw,circle,minimum size = 12pt, green!50!black]at(start4.center){};
\node[draw,circle,minimum size = 14pt, green!50!black]at(start4.center){};
\node[draw,circle,minimum size = 16pt, green!50!black]at(start4.center){};

\node[draw,circle,minimum size = 10pt,RoyalBlue]at([xshift=-1em]nw.center)(start5){};
\node[draw,circle,minimum size = 12pt, RoyalBlue]at(start5.center){};
\node[draw,circle,minimum size = 14pt, RoyalBlue]at(start5.center){};
\node[draw,circle,minimum size = 16pt, RoyalBlue]at(start5.center){};

\node[draw,circle,minimum size = 10pt]at([yshift=1.25em,xshift=-1em]nw.south)(start6){};
\node[draw,circle,minimum size = 12pt, green!50!black]at(start6.center){};
\node[draw,circle,minimum size = 14pt, green!50!black]at(start6.center){};
\node[draw,circle,minimum size = 16pt, green!50!black]at(start6.center){};
\node[draw,circle,minimum size = 10pt]at([yshift=-1.5em,xshift=0.5em]nw.north)(start7){};
\node[draw,circle,minimum size = 12pt, green!50!black]at(start7.center){};
\node[draw,circle,minimum size = 14pt, green!50!black]at(start7.center){};
\node[draw,circle,minimum size = 16pt, green!50!black]at(start7.center){};

\node[right = 3em of start3,RoyalBlue,align = center]{Network\\ Manager};

\node[anchor=south west, draw,rectangle,minimum width=4em,minimum
height=2em,align=center] at (-5,-1)(NM){Physical\\
System $i$};  
\node[draw,rectangle,minimum width=4em,minimum
height=2em,align=center,above=2em of
NM](self2){Self\\
Trigger}; 
\node[anchor=south west,draw,rectangle,minimum width=4em,minimum
height=2em,align=center,dark-gray](remote) at (-3.4,0.8){Data Source\\ $j$};
 
\node[anchor=south west,draw,rectangle,minimum width=4em,minimum
height=2em,align=center](process) at (3,-0.35){Physical\\ System $i+1$};
\node[draw,rectangle,minimum width=4em,minimum
height=2em,align=center,above=2em of
process](self){Self\\
Trigger}; 

\node[draw,rectangle,minimum width=4em,minimum height=2em,align=center,dark-gray](remote2)at(3,-1){Data Source\\  $j+1$};
\draw[->,dashed](self2.south east) -- (start3); 
\draw[->](start3) -- (NM.east);
\draw[->](NM) -- (self2);
\draw[->,dashed,dark-gray](remote.north east) -- (start2);
\draw[->,dark-gray](start2) -- (remote.south east);
\draw[->](start4) -- (process.west);
\draw[->] (process) -- (self);
\draw[->,dashed] (self.south west) -- (start4);  
\draw[->,dashed,dark-gray](remote2.north west) -- (start1);
\draw[->,dark-gray](start1) -- (remote2.south west);
\end{tikzpicture}}
\vspace{-1mm}
\caption{
We consider multiple physical systems connected over a wireless multi-hop network.
Each system is associated with a self trigger that computes at the current communication instant when it needs to communicate next.
This information is piggybacked onto the message it sends.
The network manager uses this information to compute a communication schedule respecting these demands and, if possible, reallocating bandwidth to additional data sources.
}
\label{fig:cgc}
\vspace{-5mm}
\end{figure}

At the heart of our solution is the novel concept of \emph{control-guided communication}: The control system informs the communication system \emph{at run time} about its resource requirements, and the communication system leverages this information to dynamically allocate or shut down resources.  
Concretely, we consider the setup depicted in \figref{fig:cgc}.  
Each agent uses STC to decide at the current communication instant when it will communicate next. 
The agent piggybacks the decision of its self trigger onto the messages it sends. 
The network manager uses this information as input when dynamically computing the communication schedule at run time.   
For example, when some agents do not need to communicate, their share of the bandwidth can be reallocated to serve other traffic (\eg from remote sensors) or can be shut down to conserve energy. 
The concrete scheduling policy is an exchangeable component of our design and can be adapted to the application requirements.  

In essence, we make the following two main contributions:
\begin{compactitem}
  \item We propose control-guided communication, a tight integration of STC and wireless multi-hop communication in which the control system informs the network at run time about future communication demands to enable both energy savings \emph{and} reallocation of network bandwidth.
  \item Using experiments on a real cyber-physical testbed with five inverted pendulums, we are the first to demonstrate distributed STC over wireless multi-hop networks with update intervals below \SI{100}{\milli\second}, while showing energy savings of up to \SI{87}{\percent} compared to the periodic baseline.
\end{compactitem}

\vspace{-2mm}
\section{Problem Setting}
\label{sec:problem}


We consider $N$ physical systems connected over a wireless multi-hop network, as shown in \figref{fig:cgc}.
Each agent is modeled as a stochastic, linear, and time-invariant system
\begin{align}
\label{eqn:lin_dynamics}
x_i(k+1) = A_ix_i(k)+B_iu_i(k)+v_i(k),
\end{align} 
with state $x_i(k)\in\R^{n}$, input $u_i(k)\in\R^{m}$, and $v_i(k)\in\R^n$ a Gaussian random variable with zero mean and variance $\Sigma_i$, capturing process noise.
We assume each agent has a local controller that receives local observations directly, but also needs information from other agents for distributed control.
 

There are various methods to design distributed controllers (see, for example,~\cite{Lunze1992}).
In this work, we adopt an approach based on the linear quadratic regulator (LQR)~\cite{Anderson2007}.
Using augmented states $\tilde{x}(k) = (x_1(k),\ldots,x_\mathrm{N}(k))^\transp$ and inputs $\tilde{u}(k) = (u_1(k),\ldots,u_\mathrm{N}(k))^\transp$, we define the cost function
\begin{align}
\label{eqn:LQR_cost}
J = \lim\limits_{k\to\infty}\frac{1}{K}\E[\tilde{x}^\transp (k)Q\tilde{x}(k) + \tilde{u}^\transp(k)R\tilde{u}(k)],
\end{align}
with positive definite weight matrices $Q$ and $R$. 
The optimal stabilizing controller that minimizes~\eqref{eqn:LQR_cost} is of the form $u_i(k) = \sum_j F_{ij}x_j(k)$, where $F_{ij}$ denotes entry $(i,j)$ of the feedback matrix $F$.
That is, to implement this controller, each agent needs information from all other agents, which is sent over the wireless multi-hop network.
To provide high-performance control while efficiently using limited network bandwidth and energy resources, the system must meet several requirements:
\begin{compactitem}
  \item For coordination, the agents need to exchange data; in particular, for optimal control according to~\eqref{eqn:LQR_cost}, all agents need to communicate with one another (all-to-all).
  \item Wireless multi-hop communication must be reliable and fast to support feedback control of physical systems with fast dynamics; we target mechanical systems requiring update intervals on the order of tens of milliseconds~\cite{Akerberg2011}.
  \item The network must arbitrate among multiple types of data traffic as determined by the communication schedule, while always giving highest priority to control traffic.
  \item If some fraction of the bandwidth is not allocated to any entity, this resource should be shut down to save energy.
\end{compactitem}
%
\section{Co-design Approach}
\label{sec:codesign}

The main goal of this paper is to facilitate high-performance distributed control across multi-hop wireless networks with highly adaptive resource arbitration and allocation to support multiple traffic types and save unused resources.
Prior work failed to reach this goal because the many imperfections of wireless systems, such as time-varying end-to-end delays and limited throughput, complicate the control design and make it difficult to quickly coordinate the system-wide operation and resource usage based on the current control-traffic demands.

To tackle this issue, we propose a novel co-design approach that integrates the control and communication systems in two ways.
First, the \emph{design} of the communication system tames network imperfections as much as possible, and the control system accounts for the emerging key properties and remaining imperfections.
Second, during \emph{operation}, the control system reasons about its future communication demands and informs the communication system accordingly.
The communication system, on the other hand, adapts to these demands by arbitrating the available bandwidth among different types of traffic and by shutting down resources completely to save energy when neither the control system nor any other participant needs the full bandwidth.
We call this concept \emph{control-guided communication}, which we detail in the following two sections.


In addition, our wireless communication system provides fast and reliable many-to-all communication among any set of agents, even when the agents are mobile and thereby causing the network topology to change continuously.
This feature is a key difference to traditional wireless communication systems, such as WirelessHART, and makes our co-design approach directly applicable to solve various kinds of distributed control problems that may be stated in the form of a cost function~\eqref{eqn:LQR_cost}.

%


\section{Wireless Communication System Design}
\label{sec:protocol}

We first describe the design of the wireless communication system, and detail the control design based on the emerging properties in the next section.
The wireless system builds on the periodic design in~\cite{mager2019feedback} and consists of three elements, where 2) is significantly modified and 3) is a new component:
\begin{compactenum}
 \item a \emph{hardware platform} enabling a predictable and efficient execution of all control tasks and message transfers;
 \item a \emph{multi-hop wireless protocol} that provides \mbox{many-to-all} communication with minimal, bounded end-to-end delay;
 \item an \emph{online scheduler} that dynamically assigns bandwidth to each agent based on its communication requirements.
\end{compactenum} 

\fakepar{Hardware platform}
We use a dual-processor platform~(DPP) where sensing, actuation, and control execute on an application processor (MSP432P401R, 32 bit, 48 MHz) and the wireless multi-hop protocol executes on a communication processor (CC430F5147, 16 bit, 13 MHz).
The processors communicate through the Bolt interconnect~\cite{sutton15bolt}, which provides bounded worst-case execution times for the bidirectional exchange of messages between both processors.
In this way, control and communication can efficiently execute in parallel and never interfere with each other, providing timing predictability.

\begin{figure}[t]
 \centering
 \includegraphics[width=\linewidth]{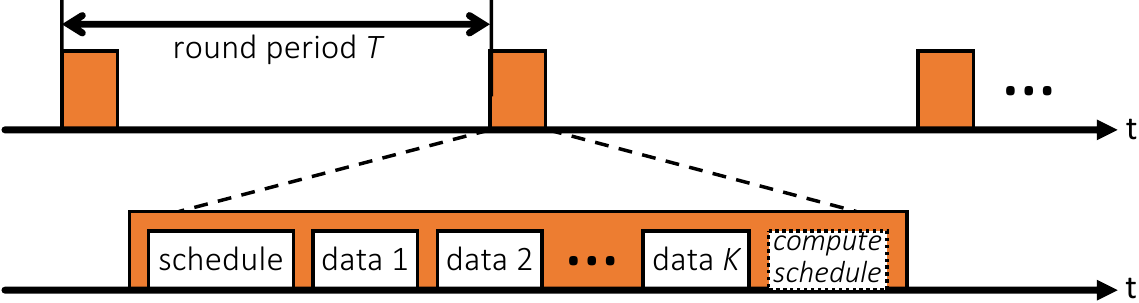}
 \vspace{-6mm}
 \caption{Time-triggered operation of the multi-hop low-power wireless protocol. Communication occurs in rounds with a constant period $T$. Each round consists of a schedule slot and up to $K$ data slots. The schedule slot serves to inform all nodes of the number of subsequent data slots in the round and the allocation of control or other messages to the scheduled data slots.}
 \label{fig:protocol}
 \vspace{-5mm}
\end{figure}

\fakepar{Multi-hop wireless protocol}
The communication processor of every \dpp in the network runs a multi-hop protocol, whose design is inspired by a new breed of protocols that exploit \emph{synchronous transmission} based flooding for highly reliable and efficient communication.
As shown in \figref{fig:protocol}, using our protocol, communication occurs in \emph{rounds} of equal duration that repeat with a constant period $T$.
Each round consists of a sequence of non-overlapping \emph{slots}.
In each slot, one node is allowed to initiate a Glossy flood~\cite{ferrari2011efficient} to send a message to all other nodes.
Glossy achieves the theoretical minimum latency for flooding a message in a multi-hop network using half-duplex radios, and provides a reliability above 99.9\,\% in real-world scenarios~\cite{ferrari2011efficient,Ferrari2012}.
In fact, Glossy's reliability can be pushed beyond 99.9999\,\% by letting nodes transmit more often during a flood, and it time-synchronizes all nodes to within sub-microsecond accuracy at no additional cost~\cite{ferrari2011efficient}.

Any node in the network can serve as the designated \emph{network manager} that uses the first slot in a round to flood the \emph{schedule}.
The schedule informs all other nodes about the number of data slots in the round (up to $K$) and the allocation of nodes to these data slots. 
The transmitted messages carry, for example, high-priority control information from agents or lower-priority data from other nodes, such as measurements from a remote sensor or information about a node's health status (\eg its battery's state of charge).
When sending a message, a node also piggybacks information about its future communication demands; if the network manager does not receive a message, it assumes that the respective node needs to transmit in the next round.
Based on all demands, the network manager computes the schedule for the next round after the last data slot.

\fakepar{Online scheduler}
To this end, the network manager maintains a list of unserved communication demands, and allocates up to $K$ nodes to the data slots in the next round according to a \emph{scheduling policy}.
The scheduling policy can be adjusted to meet different application requirements.
As an illustrative example, we design in this paper a new policy that aims to strike a balance between resource efficiency and accommodating lower-priority messages next to control traffic.
Specifically, if there are free data slots after assigning all nodes with pending control messages in the next round, we allocate one of the free data slots to a node for sending some other message (sensor, status, \etc).
The next node to send such message is chosen in a round-robin fashion.
Any other free slot is left empty.
Since nodes have their radios only on during allocated slots and off otherwise, this example policy illustrates that our wireless communication system allows for both arbitrating bandwidth among different traffic types and not allocating resources at all to save energy, as demonstrated in~\secref{sec:eval}.

\fakepar{Key properties}
Our wireless system design provides highly reliable, efficient many-to-all communication, system-wide time synchronization, and adapts at run time to the nodes' communication demands.
Due to the time synchronization, we can schedule control and communication tasks such that the jitter on the update interval and end-to-end delay is less than \SI{\pm50}{\micro\second}, as formally and experimentally validated in~\cite{mager2019feedback}.



\section{Self-Triggered Control Design}
\label{sec:ctrl}
We now detail the control design, first our approach to distributed control and then our self-triggered design.

\subsection{Distributed Control}
The wireless communication system provides a constant update interval $T$ as the jitter is negligible for the considered scenarios.
We thus set one discrete time step in~\eqref{eqn:lin_dynamics} to $T$ and data that is sent over the network is delayed by one time step.
Moreover, the many-to-all communication scheme ensures that information can be received by all agents in the network.
This greatly facilitates control design as essentially arbitrary information patterns can be implemented.
For example, this allows for implementing a (centralized) optimal controller in a distributed fashion as we show in this paper.
Given the high reliability of the wireless embedded system, we assume that data that are sent over the network are received by all agents.

As an example for distributed control, we consider synchronization of multiple agents through an LQR design as in~\eqref{eqn:LQR_cost}.
For ease of presentation, we outline the approach for the two-agent case, but it also extends to multiple agents as shown in \secref{sec:eval}.
We choose the quadratic cost function
\begin{align}
\label{eqn:cost}
J &= \lim_{K\to\infty}\frac{1}{K}\E\!\Big[\sum\limits_{k=0}^{K-1} \sum_{i=1}^2 \Big(x_i^\mathrm{T}\!(k)Q_i x_i(k) + u_i^\mathrm{T}\!(k)R_i u_i(k) \Big) \nonumber \\
&+ (x_1(k)-x_2(k))^\mathrm{T}Q_\text{sync} (x_1(k)-x_2(k)) \Big],
\end{align}
that is, we penalize deviations between $x_1(k)$ and $x_2(k)$ through the positive definite weight matrix $Q_\mathrm{sync}$, as well as deviations from the equilibrium ($Q_i>0$) and high control inputs ($R_i>0$).
Using augmented states as in~\eqref{eqn:LQR_cost}, the term in the summation over $k$ becomes
\begin{align*}
&\tilde{x}^\mathrm{T}\!(k)
\big(
\begin{smallmatrix}
Q_1+Q_\text{sync}&-Q_\text{sync}\\
-Q_\text{sync}&Q_2+Q_\text{sync}
\end{smallmatrix}
\big)
\tilde{x}(k)
 +\tilde{u}^\mathrm{T}(k)
 \big(
 \begin{smallmatrix}
R_1&0\\
0&R_2
\end{smallmatrix}
\big)
 \tilde{u}(k).
\end{align*}
As discussed in \secref{sec:problem}, solving the optimal control problem then leads to a feedback controller that has the form \mbox{$u_1(k) = F_{11} x_1(k) + F_{12} x_2(k)$}, that is, agent~$1$ needs information from agent~$2$.
We account for this by letting agent~$2$ send $u_{12}(k) = F_{12}x_2(k)$ over the network.
Thus, agent~$1$'s control input consists of $u_{11}(k)=F_{11}x_1(k)$, which it can compute using its local observations, and $u_{12}(k)$, which it receives over the network.
We can thus define the closed-loop matrix $\tilde{A}_1=A_1+B_1F_{11}$ and~\eqref{eqn:lin_dynamics} then reads as follows
\begin{align}
\label{eqn:lin_dynamics_mod}
x_1(k+1) = \tilde{A}_1x_1(k)+B_1u_{12}(k)+v_1(k).
\end{align}

\subsection{Self-triggered Approach}

Different STC designs have been proposed and are conceivable to realize control-guided communication.  
We use a design that exploits ideas from previous work on state estimation~\cite{trimpe2019resource}.  
Instead of sending states as in~\cite{trimpe2019resource}, we consider the communication of control inputs.  
Specifically, rather than sending its entire state, agent~$2$ only sends the input $u_{12}(k)$ that is needed by agent~$1$.
In case of no communication, agent~$1$ keeps applying $u_{12}(k_\ell)$, where $k_\ell$ is the last time step at which the input $u_{12}(k)$ was sent.
We trigger communication based on the error $ e_{12}(k)\coloneqq u_{12}(k)-u_{12}(k_\ell)$ as follows
\begin{align}
\label{eqn:trig_rule}
\gamma_2(k) = 1 \iff (e_{12}(k))^\transp e_{12}(k)>\delta.
\end{align}
Here, $\gamma_2(k)$ is a binary variable, denoting whether agent~$2$ communicates $u_{12}(k)$ ($\gamma_2(k)\!=\!1$) or not ($\gamma_2(k)\!=\!0$), 
while $\delta$ defines the designer's trade-off between saving communication (large $\delta$) and keeping the error to a minimum (small~$\delta$).

If we directly implement~\eqref{eqn:trig_rule}, agent~$2$ instantaneously decides on whether to transmit $u_{12}(k)$ to agent~$1$.
In case of a negative triggering decision, there is no possibility to reallocate bandwidth and hence freed resources remain unused.
To overcome this problem, we use a self-triggered strategy.
Whenever an agent communicates, it already decides when to communicate next.
To this end, we predict the evolution of the error and look for the smallest $M>1$ such that
\begin{align}
\label{eqn:self_trigger}
\E\left[(e_{12}(k+M))^\transp e_{12}(k+M)|\mathcal{D}_2(k)\right] > \delta
\end{align}
and set $\gamma(k+M-1)=1$.
Here, $\mathcal{D}_2(k)$ describes the data agent~$2$ collected until time step $k$, that is, its local states $x_2$ and the inputs $u_2$ and $u_{12}$ that it has applied and sent so far, respectively.
The rationale behind this triggering rule is as follows: Information that is sent over the network is delayed by one discrete time step.
The inequality in~\eqref{eqn:self_trigger} tells us that the error exceeds, in expectation, the threshold $\delta$ in $M$ time steps.
We thus seek to communicate next in $M-1$ time steps such that the new input arrives in $M$ time steps, which is exactly when we expect the error to exceed the threshold.

The exact computation of~\eqref{eqn:self_trigger} is complicated by the fact that the input $u_{21}(k)$ is not available at all times at agent~$2$. 
To derive the triggering law, we assume $u_{21}(k)$ is known and then comment on how we approximate it to yield a tractable implementation.
Based on this, we get the error distribution
\begin{align}
f(e_{12}(k+M)|\mathcal{D}_2(k)) = \mathcal{N}(&\hat{e}_{12}(k+M|k),P_2(k+M|k))\nonumber,
\end{align}
with mean $\hat{e}_{12}$ and variance $P_2$ given as
\begin{subequations}
\label{eqn:err_mean_var}
\begin{align}
\label{eqn:mean}
\begin{split}
&\hat{e}_{12}(k+M|k) =\\ &F_{12}(\tilde{A}_2^{M}x_2(k) + \sum_{i=0}^\mathrm{M} \tilde{A}_2^{M-i}B_2u_{21}(k+i))-u_{12}(k)
\end{split}
\end{align}
\begin{align}
&P_2(k+1|k) = F_{12}^\transp (\tilde{A}_2^\transp P_2(k|k)\tilde{A}_2+\Sigma_2)F_{12}.
\end{align}
\end{subequations}
Equations~ \eqref{eqn:err_mean_var} are standard open-loop state and covariance predictions of the system in~\eqref{eqn:lin_dynamics_mod}, so the derivations follow from Kalman filter theory~\cite[p.~111]{anderson2012optimal}.

Given this error distribution, we can now, using $\E[e^\transp e]\!=\!\norm{\E[e]}^2\!+\!\Tr(\Var[e])$, solve for the triggering rule~\eqref{eqn:trig_rule}:
At every communication instant, find the smallest $M\!>\!1$ such that
\begin{align}
\label{eqn:trigger_rule}
&\lVert F_{12}(\tilde{A}_2^{M}x_2(k) + \sum\limits_{i=0}^\mathrm{M} \tilde{A}_2^{M-i}B_2u_{21}(k+i))-u_{12}(k)\rVert^2\nonumber \\
&+ \Tr(F_{12}^\transp (\tilde{A}_2^\transp P_2(k+M|k)\tilde{A}_2+\Sigma_2)F_{12})>\delta,
\end{align}
with $\Tr$ the trace of a matrix.

So far, we assumed that agent~$2$ has knowledge about the future development of $u_{21}(k+i)$, which does not hold in practice.
Because agent~$2$ has no information about the current state of agent~$1$ and hence cannot infer the future development of $u_{21}(k+i)$, it approximates $u_{21}(k+i)$ as $u_{21}(k+i)=u_{21}(k)\,\forall i\in[0,M)$.
With this, the input $u_{21}(k+i)$ in~\eqref{eqn:mean} and~\eqref{eqn:trigger_rule} effectively becomes a constant.

We note that one way to let agent~$1$ reason about agent~$2$'s state would be to send the entire state $x_2(k)$ instead of the control input $u_{12}(k)$.
Agent~$1$ could use this state to compute $u_{12}(k)$ and to predict the evolution of agent~$2$'s state.
This, however, incurs higher communication demands at each instant as the state is typically of higher dimension than the input.

%
\vspace{-1mm}
\section{Experimental Evaluation}
\label{sec:eval}
\vspace{-0mm}

%
\begin{figure}[!tb]
    \centering
    \includegraphics[width=\linewidth]{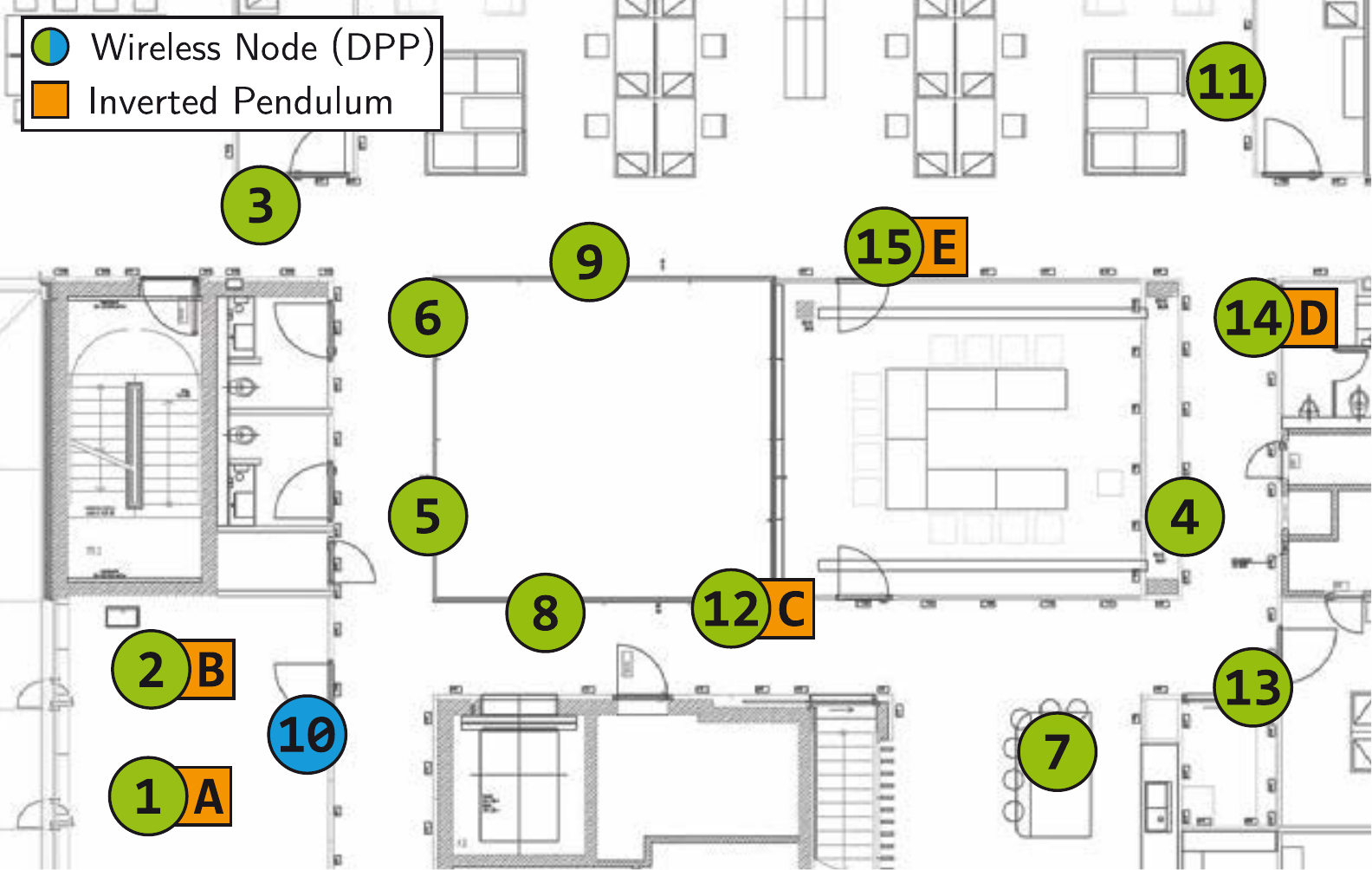}
    \vspace{-5mm}
    \caption{Cyber-physical testbed with 15 wireless DPP nodes and five cart-pole systems (A and B are real systems; C, D, and E are simulated systems). The network has a diameter of three hops. Node 10 is the network manager.}
    \label{fig:testbed}
    \vspace{-5mm}
\end{figure}

We evaluate our approach using experiments on a real cyber-physical testbed~\cite{Baumann2018, mager2019feedback} shown in \figref{fig:testbed}.
It consists of 15 wireless \dpp nodes and five cart-pole systems (or pendulums), where A and B are real systems and C, D, and E are simulated systems.
The nodes are distributed in an office space of about \SI{15}{\meter} by \SI{20}{\meter}, and transmit at \SI{-6}{\dBm} and 250\,kbit/s in the 868\,MHz band, forming a three-hop wireless network.

\vspace{-2mm}
\subsection{Scenario and Metrics}

\fakepar{Scenario}
The control task of each pendulum is to locally stabilize itself and to synchronize its cart position with all others.
Since each system has access to its local state $x_i(k)$, we can run the local feedback loop at a faster update interval than communication over the network occurs.
Here, we choose an update interval of \SI{10}{\milli\second} for the local loop.
Control inputs $u_{ij}(k)$ of the other agents are communicated over the wireless multi-hop network,
where the exchange of all control inputs takes \SI{50}{\milli\second} (\ie one communication round with up to $K=5$ data slots and 4{\kern0.15em}byte per agent).
We use the scheduling policy outlined in \secref{sec:protocol}.
To challenge the synchronization of the cart positions, we apply a sine distortion signal (\SI{3.6}{\second} period with an amplitude of \SI{\pm5}{\volt}) to the control input of pendulum~B.

The controllers are designed as described in \secref{sec:ctrl}.
We use the same model for the cart-pole system as in~\cite{mager2019feedback} and also adopt the $Q_i$ matrices used for periodic synchronization.
For $Q_\mathrm{sync}$, we set the first diagonal entry to $20$ and all other entries to zero to express our desire to synchronize the cart positions.
Further, we choose $R_i=0.01$ for all systems.

\fakepar{Metrics}
Our evaluation uses the following metrics:
\begin{compactitem}
    \item \emph{root mean square of the synchronization error (RMSE)} computed based on the cart positions of all pendulums in an experiment as a measure of control performance;
    \item \emph{utilization} of the available data slots during each round, broken down into free slots (radio off), slots used for control traffic, and slots used for additional (other) traffic;
    \item \emph{radio duty cycle}, the fraction of time a node has its radio on, which is a widely used metric in the low-power wireless networking literature (see, \eg~\cite{gnawali2009ctp,Ferrari2012}) for quantifying communication energy cost. 
\end{compactitem}

In the following, we first illustrate the run time operation of our co-designed wireless control system in a real experiment, and then evaluate the trade-off among control performance, communication energy cost, and serving additional traffic as a function of the triggering threshold.

\vspace{-1.5mm}
\subsection{Efficient Resource Arbitration and Allocation}
\vspace{-0.5mm}

\begin{figure*}[!tb]
    \centering
    \includegraphics[width=\textwidth]{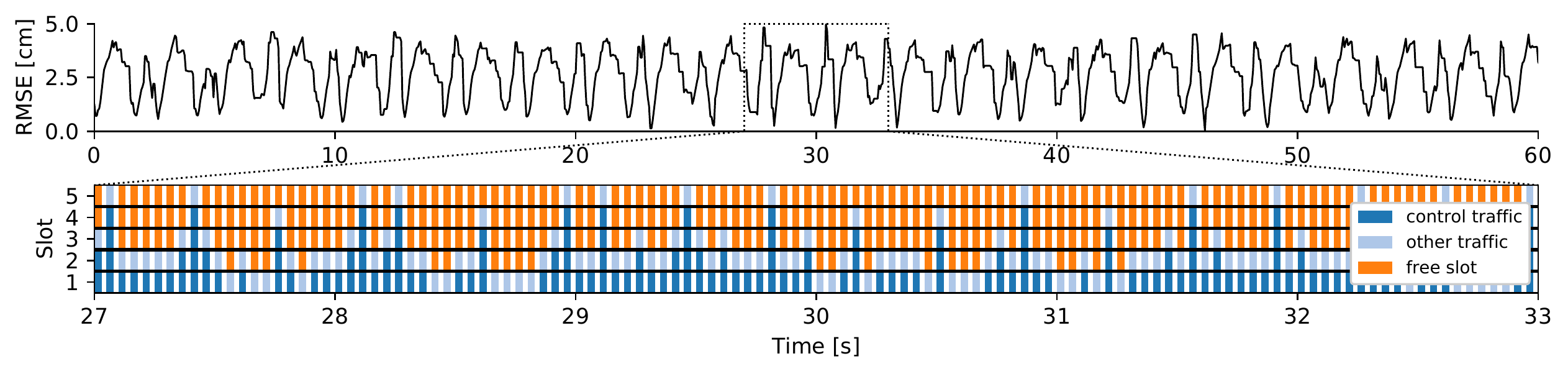}
    \vspace{-7mm}
    \caption{Control performance and bandwidth utilization over time, recorded during one of our experiments. The scheduling policy described in \secref{sec:protocol} is used but applications can also specify any other policy. Each vertical line in the lower figure represents a communication round. The control traffic demands vary over time between 0 and 4 slots. One slot is always used for other traffic and the remaining free slots are shut down to save communication energy.}
    \label{fig:slot_usage}
    \vspace{-4mm}
\end{figure*}

\figref{fig:slot_usage} shows a real trace of the control performance (top) and the slot utilization in each communication round (bottom) over time for a triggering threshold of $\delta=0.03$.
Looking at the utilization, we see that, on average, less than one third of the available bandwidth is needed for control traffic.
Our co-design approach effectively uses the freed bandwidth to schedule additional traffic (here at most one slot per round according to the example scheduling policy from \secref{sec:protocol}) \emph{and} to shut down the remaining bandwidth completely.
During the many free slots all nodes have their radios turned off, which saves significant amounts of energy.
Due to the sine distortion signal, the RMSE at the top exhibits a similar shape.

%
\begin{figure*}[!tb]
\begin{subfigure}[t]{0.32\linewidth}
    \centering
    \includegraphics[width=\textwidth]{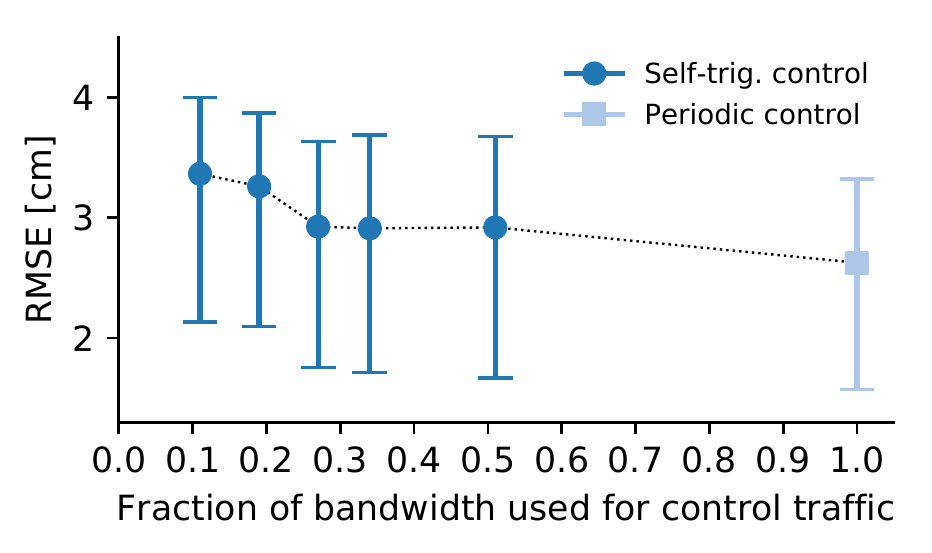}
     \vspace{-6mm}
    \subcaption{Control performance.}
    \label{fig:control_performance}
\end{subfigure}
\,
\begin{subfigure}[t]{0.32\linewidth}
    \centering
    \includegraphics[width=\textwidth]{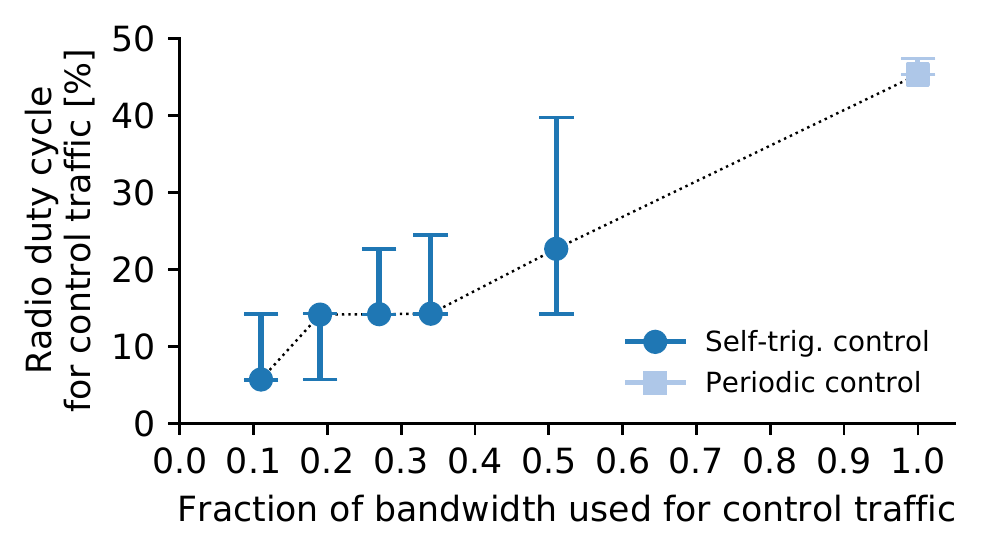}
    \vspace{-6mm}
    \subcaption{Radio-duty cycle.}
    \label{fig:rdc}
\end{subfigure}
\,
\begin{subfigure}[t]{0.32\linewidth}
    \centering
    \includegraphics[width=\textwidth]{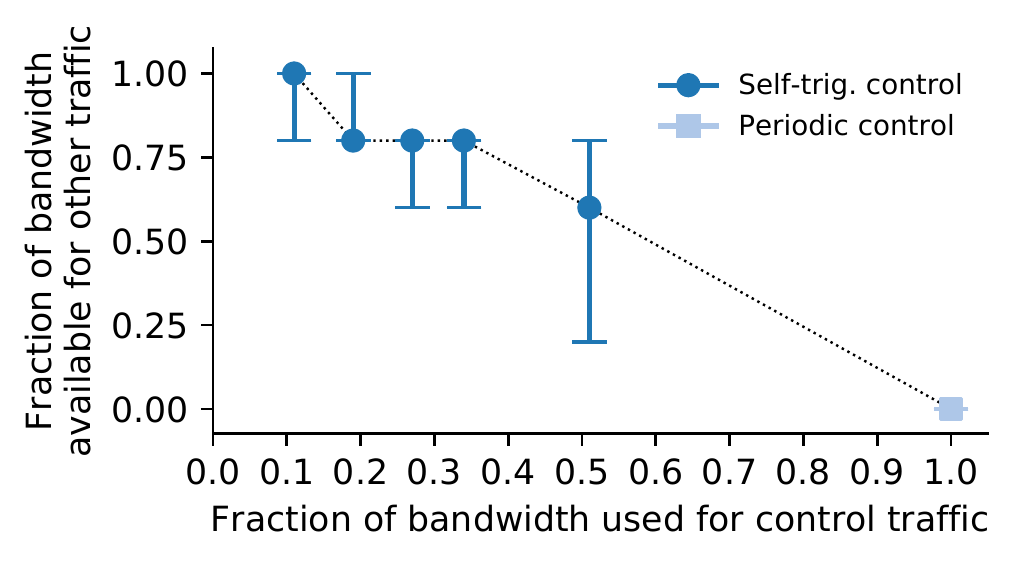}
    \vspace{-6mm}
    \subcaption{Bandwidth available for other traffic.}
    \label{fig:other_traffic}
\end{subfigure}
\vspace{-1mm}
\caption{Trade-off between control performance, communication energy efficiency, and flexibility in serving other traffic for different fractions of control traffic, reported in terms of the median and 25th/75th percentiles. Control performance decreases when less bandwidth is used for control traffic. Conversely, freed resources that are not needed for control traffic result in considerable communication energy savings or allow to serve other traffic (\eg status, sensors).}
\label{fig:performance}
\vspace{-2mm}
\end{figure*}

\vspace{-1.5mm}
\subsection{Control Performance vs.~Efficiency vs.~Flexibility}
\vspace{-0.5mm}

The triggering threshold $\delta$ allows a user to trade control performance for communication energy efficiency and flexibility in serving other traffic.
To evaluate this trade-off, we consider six different thresholds and perform for each threshold three 2-minute experiments.
In addition, we perform experiments with $\delta = 0$ to obtain results for periodic control, where all agents communicate in every time step requiring all bandwidth for control traffic.
For each threshold, we report the median and 25th/75th percentiles across the three experiments.

\figref{fig:performance} shows RMSE, radio duty cycle for control traffic, and fraction of bandwidth available for other traffic against the fraction of bandwidth used for control traffic.
We use this intuitive unit for the x-axis instead of the triggering threshold $\delta$ because our measurements reveal that each $\delta$ corresponds to a certain fraction of bandwidth used for control traffic with negligible variance across experiments with the same $\delta$.

Looking at \figref{fig:performance}, we observe that the more bandwidth is used for control traffic, the better the control performance and the less bandwidth is available for other traffic.
As expected, higher bandwidth demands result in a higher radio duty cycle.
Using \SI{25}{\percent} of the available bandwidth for control traffic, the control performance is still comparable to the periodic baseline.
Further bandwidth reductions lead to a noticeable decrease in control performance compared with the periodic baseline of up to \SI{22}{\percent} when only \SI{11}{\percent} of the available bandwidth is used for control traffic.
At the same time, up to \SI{87}{\percent} of communication energy can be saved, while the vast majority of the bandwidth is available for other traffic.
Overall, these experimental results demonstrate that our control-guided communication approach allows for exploiting this trade-off to meet a wide range of requirements of emerging cyber-physical applications.

\vspace{-0mm}
\section{Conclusions}
\label{sec:conclusion}
\vspace{-1mm}

We have demonstrated for the first time distributed, self-triggered control over wireless multi-hop networks with energy savings \emph{and} reallocation of resources at fast update intervals of tens of milliseconds.
At the heart of our solution is control-guided communication, a new co-design approach where the control system predicts and informs the communication system about future resource demands.
Using this information, bandwidth and energy are either saved or used efficiently for different kinds of traffic.
Experiments on a real cyber-physical testbed show the effectiveness of our approach.
As part of our future work, we focus on a variety of theoretical questions, for example, regarding the closed-loop stability of the overall system, especially in the presence of message loss.

\vspace{-1mm}

\section*{Acknowledgements}
We thank Harsoveet Singh and  Felix Grimminger for
help with the testbed, and the TEC group at ETH Zurich
for making the design of the DPP platform available to the
public.


\bibliographystyle{IEEEtran}
\bibliography{IEEEabrv,refs}

\end{document}